 \documentclass[twocolumn,trackchanges,twocolappendix]{aastex631}

\usepackage{hyperref}
\usepackage{amsmath, amssymb}
\usepackage{float}

\begin{document}

\title{Relativistic Scattering in the Funnel of Cygnus X-3 }

\author[0009-0004-9795-9820]{Suraj K. Chaurasia}
\affiliation{Department of Physics, Institute of Science, Banaras Hindu University, Varanasi-221005, India}

\author{Ranjeev Misra}
\affiliation{Inter-University Centre for Astronomy and Astrophysics (IUCAA), PB No.4, Ganeshkhind, Pune-411007, India}

\author{Amit Pathak}

\affiliation{Department of Physics, Institute of Science, Banaras Hindu University, Varanasi-221005, India}



\begin{abstract}

Cygnus X-3 presents significant challenges to standard accretion models. Recent polarimetric observations by IXPE reveal high polarization degrees (PD) in the hard state ($\sim 23\%$) and unexpectedly significant polarization in the soft state ($\sim 12\%$), which are difficult to reconcile with static scattering models at low inclination ($i \approx 30^{\circ}$). We present a relativistic scattering model within a funnel-shaped geometry that resolves this discrepancy. We show that a single funnel-outflow configuration with variable bulk velocity $\beta$ can reproduce both polarization states, with lower velocities ($\beta \approx 0$) yielding $\sim 12\%$ polarization (soft state) and mildly relativistic velocities ($\beta \lesssim 0.4$) producing $\sim 23\%$ polarization (hard state) at $i \approx 30^{\circ}$ for half funnel opening angles of $\sim 13^{\circ}$–$16^{\circ}$. Relativistic aberration modifies the effective scattering angle in the comoving frame, enhancing polarization in the hard state while recovering the static limit in the soft state. The model also yields a consistent estimate of the intrinsic luminosity, of order $\sim 10^{40}\ \mathrm{erg\ s^{-1}}$, supporting a super-Eddington interpretation. This framework provides a unified explanation of the observed polarization properties of Cygnus X-3.


\end{abstract}


\section{Introduction} 

Cygnus X-3 stands as one of the most luminous and persistent radio emitting X-ray binaries in the Galaxy, serving as a critical laboratory for understanding extreme accretion physics. Discovered in 1966 (\citealt{1967ApJ...148L.119G}), Cygnus X-3 has long puzzled astronomers due to its unusual properties, including a short 4.8-hour orbital period (\citealt{1986ApJ...309..700M}), a Wolf–Rayet star companion (\citealt{vanKerkwijk1992}), and giant radio flares linked to relativistic jets. Located at a distance of $9.67 \pm 0.5$ kpc (\citealt{Reid_2023}) in the Cygnus OB2 association, the system is heavily obscured by the Galactic plane, rendering it invisible in optical bands and necessitating multi-wavelength observations in X-ray, infrared, and radio regimes. The inclination of Cygnus X-3 is constrained to be approximately $30^\circ$, based on photoionization simulations and studies of the orbital modulation of emission lines (\citealt{refId2}). Additionally, \citealt{Antokhin_2022} derived an inclination of $29.5^\circ \pm 1.2^\circ$ through analysis of X-ray and infrared light curves. The nature of the compact object in Cygnus X-3 remains a subject of ongoing debate. While its spectral properties closely resemble those of black hole X-ray binaries favoring a black hole interpretation (\citealt{10.1111/j.1365-2966.2008.13479.x}), the possibility of a neutron star accretor cannot be definitively excluded. Nevertheless, the system’s intrinsic X-ray luminosity frequently exceeds the Eddington limit, placing it firmly within the ultraluminous X-ray source (ULX) regime.

Constraining the geometry and kinematics of the accretion flow in Cygnus X-3 is therefore crucial for advancing our understanding of super-Eddington accretion and the mechanisms driving powerful outflows and jet formation. \\
The recent advent of X-ray polarimetry has provided a new dimension of diagnostic power. Polarization is uniquely
sensitive to the geometry of the scattering material, breaking the degeneracies inherent in spectral and timing analysis.
The IXPE mission has reported critical constraints for Cyg X-3 that have precipitated a crisis in current modeling
efforts. In the radio-loud hard state, the system exhibits an exceptionally high polarization degree (PD) of
approximately 23\% (\citealt{Veledina2024}), with the electric vector position angle (EVPA) generally orthogonal to the
radio jet axis. In the radio-quiet soft/ultrasoft state, the polarization decreases but remains substantial at
approximately 12\% (\citealt{refId9}). These values are anomalous because
independent constraints on the binary orbit derived from the modulation of X-ray and infrared flux fix the system
inclination at a low angle of $i \approx 30^{\circ}$. In standard accretion disk models, X-ray polarization arises from
Thomson scattering in the disk atmosphere. For a face-on or low inclination disk ($i < 45^{\circ}$), the symmetry of the
system results in near-cancellation of the polarized flux, predicting a PD of less than 1–2\% (\citealt{Schnittman_2010}). The observation of 12–23\%
polarization at $30^{\circ}$ inclination implies that the scattering geometry is not a flat disk but rather an optically
thick, geometrically elevated structure that breaks the viewing symmetry specifically, a funnel or cone.
\\
To address these high PD values, recent studies have used a static funnel model (\citealt{Veledina2024,
refId9, Chaurasia_2025}). In this scenario, the compact object is obscured by a funnel structure with a narrow opening angle ($< 15^{\circ}$). While this geometry represents an improvement over the flat disk, calculations utilizing static radiative transfer fail to provide a unified
solution. A model assuming reflection from the funnel walls can reproduce the 23\% PD of the hard state but overpredicts
the polarization in the soft state. 
A model with volume scattering within the funnel gas
can match the 12\% PD of the soft state. However, it fundamentally fails to reach the 23\% level required for the hard
state. At $30^{\circ}$ inclination, the maximum polarization achievable by static Thomson scattering in a cone is
geometrically limited to $\sim 14\%$. Consequently, current literature has been forced to invoke different physical
mechanisms (switching between reflection dominated and scattering dominated geometries) to explain the different
spectral states. This ad-hoc approach contradicts the observational evidence of a persistent global geometry and lacks a unifying physical driver. The critical role of relativistic motion in interpreting X-ray polarimetric data has recently been demonstrated for the prototypical black hole binary Cygnus X-1 (\citealt{Poutanen_2023}). Recent observations by the Imaging X-ray Polarimetry Explorer (IXPE) revealed a polarization degree of approximately $4\%$, which significantly exceeds the $\le1\%$ predicted by standard static models at the system's low orbital inclination of $i \approx 30^{\circ}$. \cite{Poutanen_2023} successfully resolved this discrepancy by proposing a model where the X-ray emitting plasma outflows with a mildly relativistic terminal velocity of $\beta_{0} \ge 0.4$. In this framework, relativistic aberration modifies the angular distribution of seed photons in the plasma rest frame, naturally enhancing the polarization of the escaping radiation at low inclinations. This finding underscores that the dynamic state of the scattering medium is a fundamental ingredient in explaining the high polarization observed in microquasars, providing a strong precedent for the relativistic funnel framework applied here to Cygnus X-3. \\
 This report proposes a resolution to the polarization puzzle by relaxing the assumption of
distinct static geometries. We introduce a Relativistic Scattering Model, where the material within the funnel is moving outward along the funnel axis with a relativistic velocity $\beta = v/c$. \\
Cyg X-3 is a known microquasar capable of launching jets with $\beta > 0.5$ (\citealt{Mioduszewski_2001, Miller-Jones_2004}). By incorporating relativistic motion, we introduce two transformative effects, the energy and flux of photons are boosted in the direction of motion, altering the effective brightness distribution of the funnel medium and the angle of scattering is transformed between the laboratory frame and the plasma rest frame. Aberration can shift the effective scattering angle closer to $90^{\circ}$ (where Thomson polarization is maximized), even when the geometric viewing angle is fixed at $30^{\circ}$. \\
We demonstrate that a single model geometry of a funnel with variable outflow velocity can naturally explain both the 12\% PD (corresponding to a slow/static wind in the soft state) and the 23\% PD (corresponding to a relativistic wind in the hard state). \\
\section{The relativistic funnel model}
\begin{figure}
    \centering
    \includegraphics[width=0.8\linewidth]{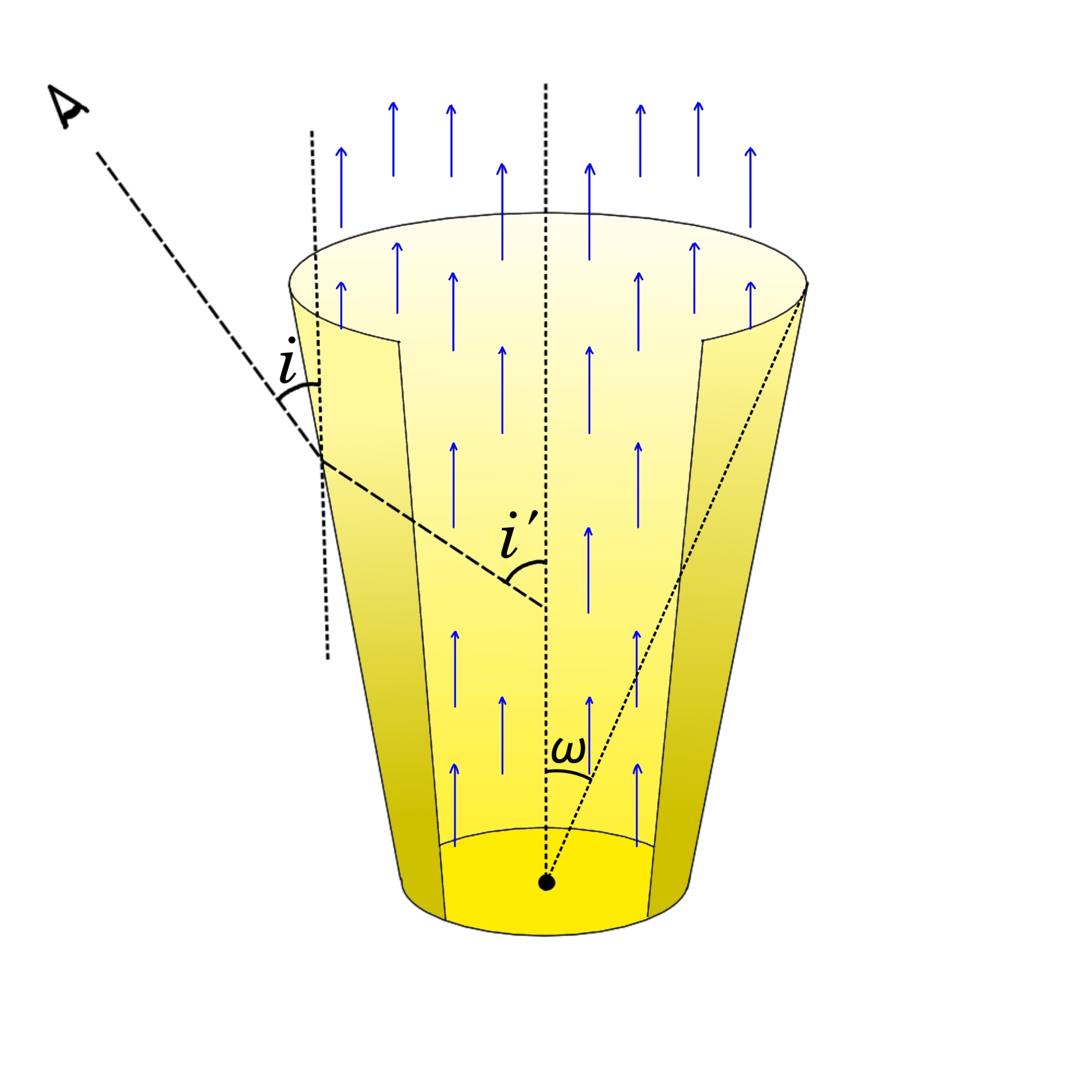}
    \caption{Schematic representation of the relativistic funnel geometry with a central primary X-ray source (black dot) located at the origin, obscured by the surrounding optically thick medium. The interior scattering medium propagates upward (blue arrows) with a macroscopic bulk velocity $v$ (where $\beta = v/c$). The dashed line indicates the observer's line of sight, forming an inclination angle $i$ and $i'$ relative to the symmetry axis of the funnel in lab frame and comoving frame respectively.}
    \label{fig:fig1}
\end{figure}

The proposed relativistic scattering model is founded upon a funnel-shaped geometry that replaces the traditional flat accretion disk assumption, as illustrated in Figure \ref{fig:fig1}. In this configuration, a central primary X-ray source is located at the origin and is effectively obscured by surrounding optically thick medium that form a geometrically elevated structure. Within the interior of this funnel, a scattering medium which may represent a jet sheath or entrained disk wind propagates radially outward along the symmetry axis with a macroscopic bulk velocity $v$, characterized by the dimensionless parameter $\beta=v/c$. The observer views the system at a fixed laboratory frame inclination $i \approx 30^{\circ}$, but due to the high-velocity motion of the scatterers, relativistic aberration transforms the direction of light rays such that the effective scattering angle in the comoving frame, $\Psi^{\prime}$, can shift toward higher values. This geometric framework, defined by a narrow half-opening angle $\omega$ typically between $13^{\circ}$ and $16^{\circ}$, allows a single physical configuration to naturally reproduce the observed polarization states of Cygnus X-3 by varying only the outflow velocity $\beta$.

\section{Theoretical Framework: The Relativistic Scattering Model}
\label{sec:theoretical framework}

\begin{figure}[h]
    \centering
    \includegraphics[width=1\linewidth]{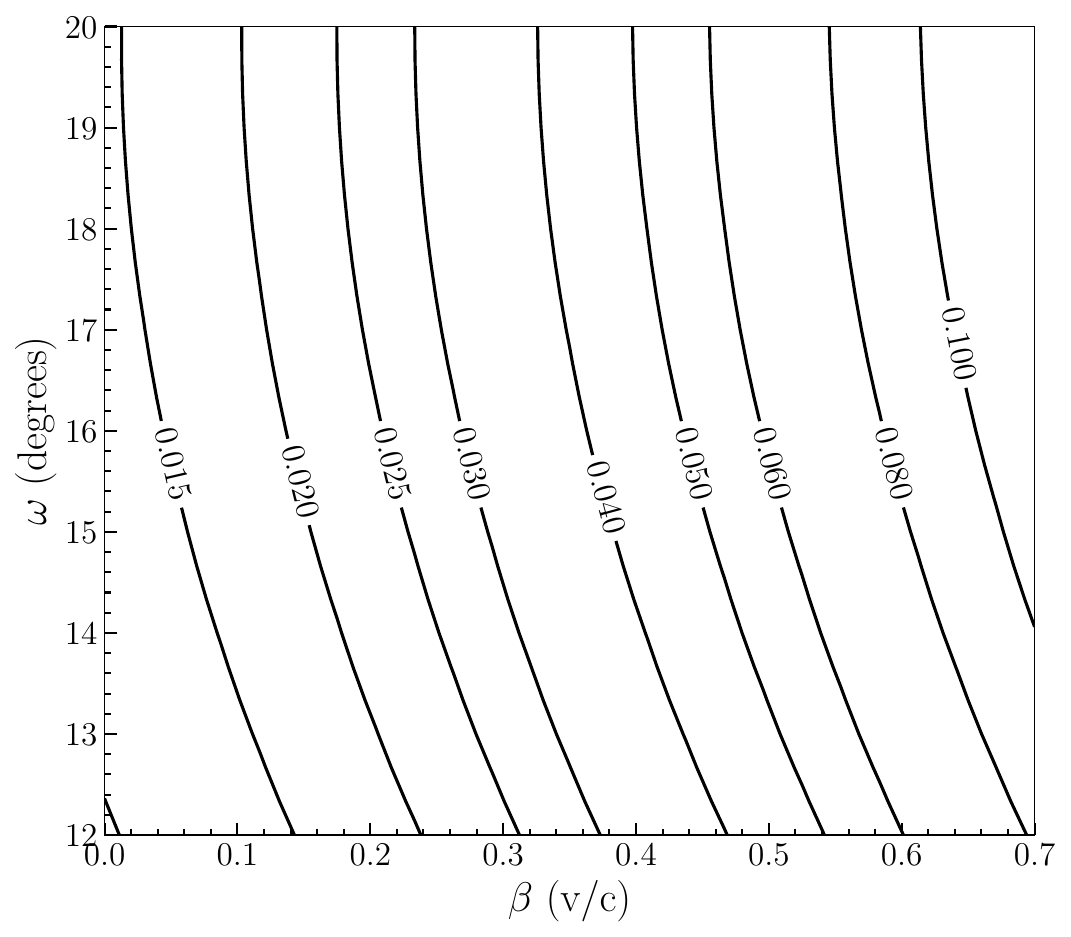}
    \includegraphics[width=1\linewidth]{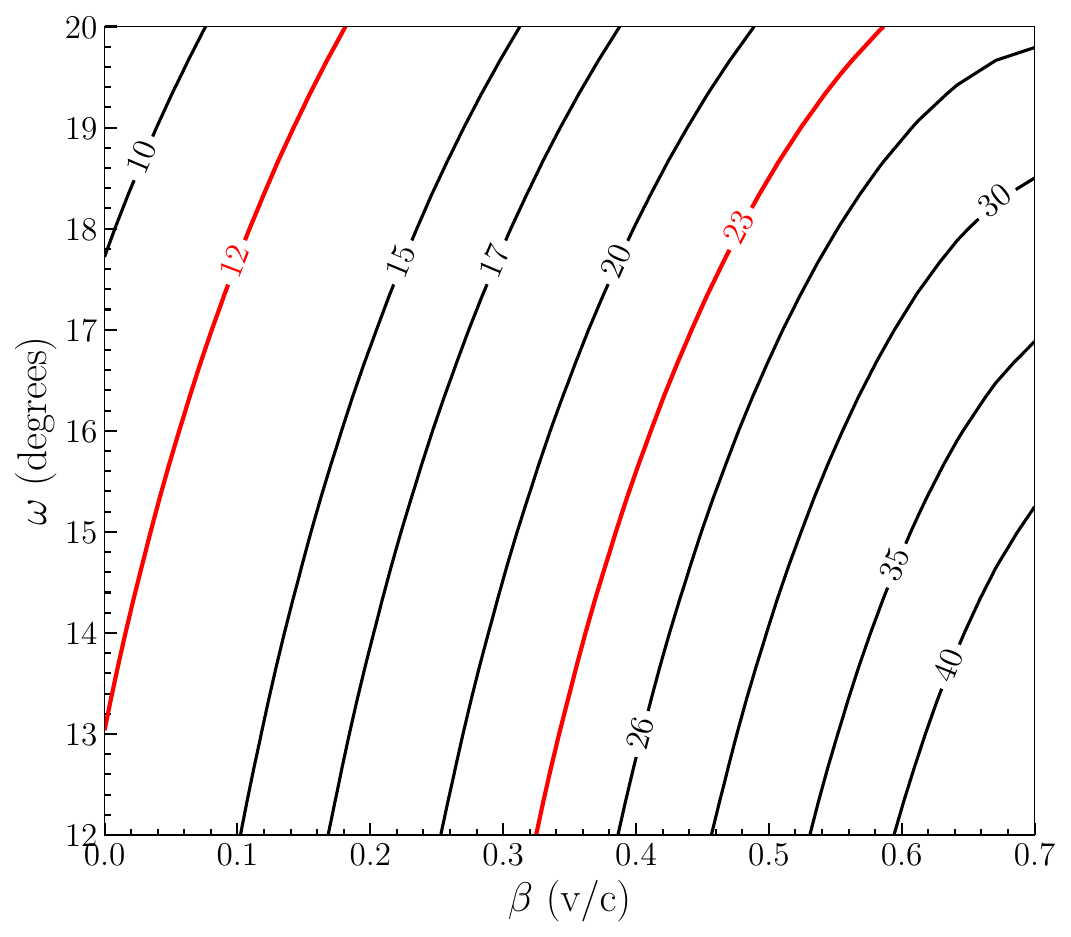}
    \caption{Top: Contours of constant flux (in units of $\lambda F_{i}$). Bottom: Contours of constant polarization degree (PD), for radiation undergoing single scattering within a funnel of half-opening angle $\omega$ and bulk velocity parameter $\beta = v/c$, computed at a fixed observer inclination of $i = 30^\circ$.}
    \label{fig:fig2}
\end{figure}

\begin{figure}[h]
    \centering
    \includegraphics[width=1\linewidth]{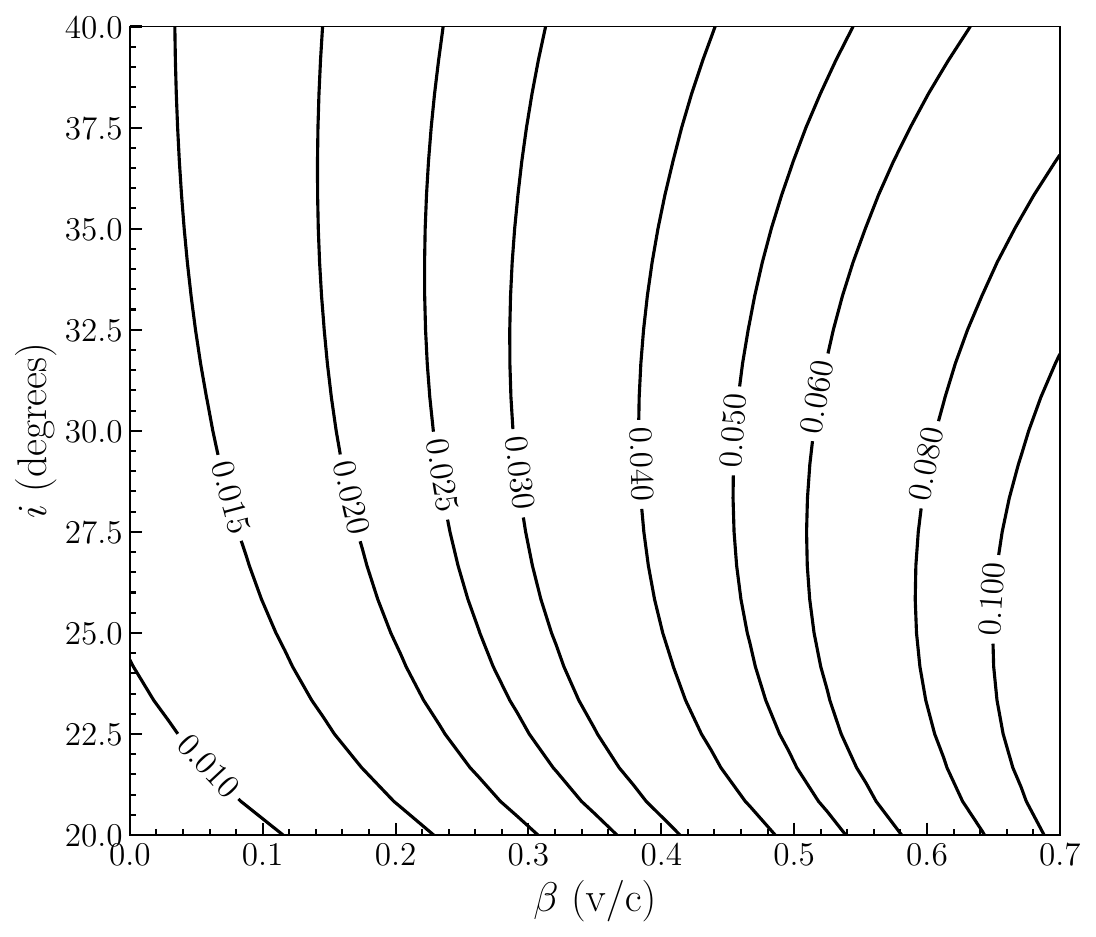}
    \includegraphics[width=1\linewidth]{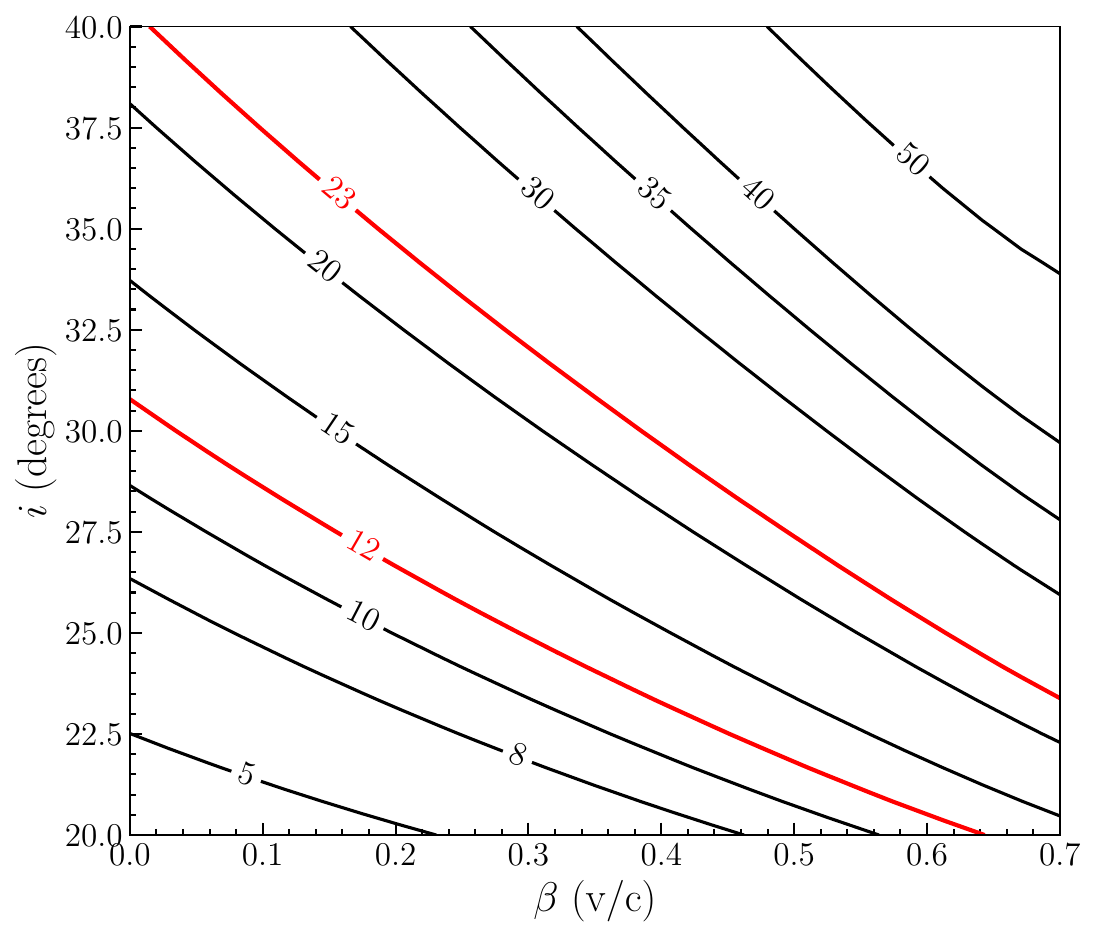}
    \caption{Top: Contours of constant flux (in units of $\lambda F_{i}$). Bottom: Contours of constant polarization degree (PD) for radiation undergoing single scattering within a funnel with bulk velocity parameter $\beta = v/c$ and  observer inclination $i$, for a fixed half-opening angle $\omega = 15^\circ$.}
    \label{fig:fig3}
\end{figure}

To construct the relativistic scattering model, we transform the radiative transfer equations to account for the macroscopic motion of the scattering medium. By transitioning from a static Euclidean framework to a Special Relativistic treatment, we define two distinct inertial frames of reference to describe the interaction between the radiation field and the high-velocity plasma.

The laboratory frame ($K$) corresponds to the rest frame of the observer and the central compact source. In this frame, the scattering medium moves with a bulk velocity $v$, and all physical quantities are denoted without primes (e.g., $\nu, \Psi, \Omega$). The comoving frame ($K^{\prime}$) is the rest frame of the plasma element involved in the scattering process. In this frame, the scattering electrons are assumed to possess a thermal, isotropic distribution. Quantities measured in this frame are indicated with primes (e.g., $\nu^{\prime}, \Psi^{\prime}, \Omega^{\prime}$). The normalized velocity of the flow is defined as $\beta = v/c$. We assume that the motion of the scattering medium is directed radially outward along the funnel axis, with a corresponding Lorentz factor $\gamma = (1-\beta^{2})^{-1/2}$. The geometry of the system, including the bulk motion of the scattering medium and the transformation of observer angles between the laboratory and comoving frames, is illustrated schematically in Figure~\ref{fig:fig1}.

\subsection{Doppler Boosting and Relativistic Aberration}

The transformation of radiation fields between the $K$ and $K^{\prime}$ frames is governed by the Doppler factor ($D$). To fully describe the complete scattering process, we must define two distinct Doppler factors: one for the photon incident on the electron ($D_{in}$) and one for the photon after scattering ($D_{out}$). 

Let $\nu_{in}$ be the photon frequency emitted by the central source in the laboratory frame, and $\nu^{\prime}$ be the corresponding frequency observed in the comoving frame of the electron. The incoming Doppler factor characterizes how highly anisotropic radiation from the central source is perceived in the rest frame of the moving electron:
\begin{equation}
D_{in} = \frac{\nu^{\prime}}{\nu_{in}} = \frac{1}{\gamma(1 - \beta \cos \phi^{\prime})}
\end{equation}
where $\phi^{\prime}$ is the angle between the incident photon propagation direction and the velocity vector of the flow in the comoving frame.

The outgoing Doppler factor describes how the scattered radiation, emitted in the comoving frame, is perceived by the distant observer in the laboratory frame:
\begin{equation}
D_{out} = \frac{\nu_{obs}}{\nu^{\prime}} = \frac{1}{\gamma(1 - \beta \cos i)}
\end{equation}
where $\nu_{obs}$ is the photon frequency measured by the distant observer, and $i$ is the inclination angle between the observer's line of sight and the funnel axis in the laboratory frame.

Crucially, the direction of the light rays is altered by relative motion. The geometric scattering angle $\Psi$ in the laboratory frame transforms to $\Psi^{\prime}$ in the comoving frame via the relativistic aberration formula:
\begin{equation}
\cos \Psi^{\prime} = \frac{\cos \Psi - \beta}{1 - \beta \cos \Psi}
\label{eq:aberration}
\end{equation}
Equation \ref{eq:aberration} is the pivotal mechanism of our model. It dictates that a photon scattered at a geometric viewing angle of $\Psi = 30^{\circ}$ in the laboratory frame corresponds to a vastly different scattering angle $\Psi^{\prime}$ in the electron's rest frame. Because the fractional polarization of Thomson-scattered radiation is entirely dependent on $\Psi^{\prime}$, this kinematic frame shift grants access to high-polarization regimes ($\Psi^{\prime} \approx 90^{\circ}$) that are strictly geometrically inaccessible in a static laboratory frame. A full volumetric derivation, incorporating relativistic effects and utilizing the Lorentz invariance of the ratio of specific intensity to the cube of frequency ($I_{\nu}/\nu^{3}$), is presented in Appendix~\ref{app:app1}. \\
The contour plots of constant scattered flux (in units of $\lambda F_{i,\mathrm{rel}}$) and constant polarization degree (PD) as a function of velocity and funnel half-opening angle ($\omega$), for a fixed radial optical depth at the base of the funnel ($\tau_{\rho} = 0.01$), observer inclination of $30^\circ$, and vertical optical depth boundaries corresponding to the obscured base ($\tau_{z,\min} = 0.1$) and maximum extent of the funnel ($\tau_{z,\max} = 20$), are shown in Figure~\ref{fig:fig2}. Here, $\lambda$ denotes the single-scattering albedo (the ratio of scattering to total extinction efficiency), and $F_{i,\mathrm{rel}}$ represents the intrinsic source flux in the relativistic scattering framework.
The contour plots of constant scattered flux (in units of $\lambda F_{i,\mathrm{rel}}$) and PD, for scattering within the funnel volume, are shown in Figure~\ref{fig:fig3} for different observer inclinations and velocities at fixed $\omega$, with the same vertical optical depth boundaries ($\tau_{z,\min} = 0.1$, $\tau_{z,\max} = 20$). \\

\section*{Luminosity Scaling in the Relativistic Scattering Regime}The determination of the intrinsic X-ray luminosity of Cygnus X-3 depends fundamentally on the efficiency of the scattering geometry. In previous investigations utilizing a static funnel approach, the high polarization degree of 23\% observed in the hard state could only be reproduced via a reflection-dominated scenario with a highly restricted flux ratio. We contrast this with our new relativistic scattering model. 
The relationship between the observed unabsorbed flux $F_{obs}$ and the intrinsic flux $F_i$ is mediated by the scattered flux ratio $R_{model}$:
\begin{equation}
F_{obs} = R_{model} \cdot \lambda \cdot F_i
\end{equation}
Using the unabsorbed flux $2.56 \times 10^{-7} \text{ erg s}^{-1} \text{ cm}^{-2}$ and distance $d = 9.7 \text{ kpc}$, the previous spectral fit established a baseline plane disk luminosity $L' = 2.88 \times 10^{39} \text{ erg s}^{-1}$ with a flux ratio $R_d = 0.323$ (\citealt{Chaurasia_2025}). In the current relativistic framework, we define a scaling factor $\eta = R_d / R_{f,rel}$, where $R_{f,rel}$ is extracted from the volumetric integration of the relativistic scattering master equation. For the soft and hard spectral states, velocities of $\beta = 0.02$ and $\beta = 0.36$ are required to reproduce polarization degrees of $12\%$ and $23\%$, respectively. We therefore evaluate these cases at a fixed inclination of $i = 30^\circ$.

In the static case, where the observed flux is determined purely by geometry, the flux ratio $R_{f,r}$ is $\sim 1.77 \times 10^{-3}$, reflecting the intrinsic inefficiency of reflection from a narrow funnel at low inclinations. 

In contrast, in our relativistic model, the inclusion of the Doppler boosting significantly enhances the observed flux, increasing the ratio to $\sim 0.0122$ and $\sim 0.0347$ for the soft and hard states, respectively. 

Consequently, the inferred intrinsic luminosities are:
\begin{equation}
L_{\mathrm{rel}} = \eta \, L' \approx 7.62 \times 10^{40} \, \mathrm{erg\,s^{-1}} \quad \text{(soft state)},
\end{equation}
\begin{equation}
L_{\mathrm{rel}} = \eta \, L' \approx 2.68 \times 10^{40} \, \mathrm{erg\,s^{-1}} \quad \text{(hard state)}.
\end{equation}

For comparison, the static reflection model yields an intrinsic luminosity of $L \approx 5.3 \times 10^{41} \, \mathrm{erg\,s^{-1}}$ for the hard state. Our relativistic treatment therefore reduces the required luminosity by a factor of $\sim 20$, corresponding to approximately one order of magnitude. We note that, a single representative observed flux has been adopted for both the soft and hard states in the present calculation. While the actual flux may vary between states, this assumption allows us to isolate and highlight the impact of relativistic effects on the inferred intrinsic luminosity.

\subsection{Relativistic Polarimetry in the Narrow-Funnel Limit}

To isolate the kinematic effects and derive an analytical understanding of the polarization enhancement, we consider the idealized limit of an infinitely narrow cylindrical funnel. In this geometry, the wind propagates outward along the $z$-axis ($v \parallel z$), and the seed photons are emitted radially from the central source. The incident photon direction is therefore parallel to the velocity of the flow. After transforming to the comoving frame, the photon continues to propagate along the symmetry axis, implying an incoming angle of $\phi^{\prime} = 0^{\circ}$.

The observer is located at an inclination angle $i$ with respect to the flow axis in the laboratory frame. Due to relativistic aberration, the direction of the scattered photon in the comoving frame differs from the laboratory frame, transforming as:
\begin{equation}
\cos \Psi^{\prime} = \frac{\cos i - \beta}{1 - \beta \cos i}
\end{equation}

In the rest frame of the scattering electron ($K^{\prime}$), the polarization properties of Thomson-scattered radiation depend solely on this scattering angle $\Psi^{\prime}$. The resulting intrinsic polarization degree is:
\begin{equation}
PD^{\prime} = \frac{1 - \cos^{2}\Psi^{\prime}}{1 + \cos^{2}\Psi^{\prime}}
\end{equation}

For a single photon trajectory in the X-ray regime, the polarization degree is Lorentz invariant. Consequently, the observed polarization degree in the laboratory frame ($PD_{obs}$) is approximately equal to $PD^{\prime}$. Substituting the aberration relation into the polarization formula yields the equation for the relativistic scattering model:
\begin{equation}
PD_{obs}(\beta, i) = \frac{1 - \left(\frac{\cos i - \beta}{1 - \beta \cos i}\right)^{2}}{1 + \left(\frac{\cos i - \beta}{1 - \beta \cos i}\right)^{2}}
\label{eq:pd_rel}
\end{equation}

This expression generalizes the static Thomson-scattering result. In this dynamic framework, the polarization degree is no longer determined solely by the viewing inclination $i$, but is fundamentally coupled to the bulk velocity $\beta$ of the scattering medium. Increasing the bulk velocity to $\beta \approx 0.24$ raises the polarization degree from $\sim 14\%$ to $\sim 23\%$ at an observer inclination of $30^\circ$ (Appendix~\ref{app:app3}).

\section{Discussion}

The IXPE polarimetric observations of Cygnus X-3 suggest a geometry in which the central X-ray source is obscured by a geometrically thick, narrow funnel. The currently available data indicate a polarization degree (PD) of $\sim 23\%$ in the hard state and $\sim 12\%$ in the soft state, accompanied by a nearly constant polarization angle (PA $\approx 90^\circ$). Such characteristics can be naturally explained if the observed emission is dominated by scattered, rather than direct emission from the central source.

In this paper, we provide an explanation using a relativistic scattering model that offers a unified framework for interpreting the polarization behavior. In this picture, the same narrow funnel geometry can produce different polarization degrees in different spectral states through changes in the bulk velocity of the outflowing plasma. When scatterers move at relativistic speeds, relativistic aberration enhances the polarization signal. This effect has recently been invoked to explain the IXPE observations of Cygnus X-1 (\citealt{Poutanen_2023}). Within our funnel geometry, a fast outflow during the hard state would Doppler-focus radiation toward the observer at a larger effective scattering angle, increasing the polarization degree to $\sim23\%$. As the outflow velocity decreases during the soft state, the model smoothly transitions to the static scattering limit, producing polarization levels of $\sim10$--$12\%$. In this scenario, the observed polarimetric duality may be directly linked to changes in the accretion rate and higher mass accretion rates can drive faster winds and therefore produce higher polarization degrees.
The relativistic scattering model resolves the conflict inherent in static analyses. We no longer need to postulate that the geometry changes from a ``scattering cloud" to a ``reflecting wall" to explain the PD jump from 12\% to 23\%. Outflow Velocity $\beta$ aligns with the known physics of state transitions. The transition from Soft to Hard state is defined by the emergence of the non-thermal jet. It is physically inevitable that the scattering medium surrounding the jet would be entrained and accelerated, increasing $\beta$ and thus boosting the polarization via aberration.

Despite the success of the present model in reproducing the observed polarization properties, several limitations should be acknowledged. The current calculations are performed under the assumption of single scattering. While this approximation is adequate in the soft X-ray regime considered here, it may require revision if future observations extend polarization measurements to higher energies where relativistic effects become important. In addition, the inclusion of multiple scattering, as well as a more realistic treatment of the wind structure, could modify the predicted polarization degree, potentially leading to lower values.

Furthermore, the funnel geometry adopted in this work represents an idealized configuration. In realistic systems, the presence of a dense Wolf–Rayet stellar wind, along with possible jet or wind-driven outflows, is expected to produce a complex and highly structured environment around the compact object. Incorporating such effects will be important for developing a more comprehensive description of the system.

  \section{Conclusion}

This work presents a relativistic scattering model for Cygnus X-3 that unifies the complex spectral and polarimetric behavior observed by IXPE. By replacing the inadequate static funnel geometry with a dynamic relativistic framework, we demonstrate that Doppler boosting factors ($D_{\mathrm{in}}, D_{\mathrm{out}}$) and relativistic aberration are essential ingredients that significantly modify the observed polarization properties. In this framework, the observed polarization states are naturally explained as follows:
\begin{itemize}
    \item Velocities in the range $0 \leq \beta \lesssim 0.4$ can reproduce both polarization states at a fixed inclination ($i \approx 30^{\circ}$).
    \item The observed PD is consistent with funnel half-opening angles in the range $\sim 13^{\circ}$-$16^{\circ}$.
    
    \item The inferred intrinsic luminosity in both states is of order $\sim 10^{40}\ \mathrm{erg\ s^{-1}}$, supporting a super-Eddington regime.
    \item The variation in polarization degree acts as a diagnostic of the bulk outflow velocity.
    \item No change in the underlying geometry is required; the observed differences arise primarily due to relativistic effects.
\end{itemize}

A single, self-consistent model therefore eliminates the need for ad hoc geometric transitions and resolves the luminosity inconsistencies inherent in static reflection scenarios. These results strongly support the interpretation of Cygnus X-3 as a super-Eddington accretor observed along the axis of a relativistic funnel.

\section*{acknowledgments}
SKC acknowledges the support from UGC, Government of India for providing fellowship under the UGC-JRF scheme (NTA Ref. No.: 201610266337), and Inter-University Centre for Astronomy and Astrophysics (IUCAA), Pune for the IUCAA Visitors Programme. AP acknowledges financial support from the IoE grant of Banaras Hindu University (R/Dev/D/IoE/Incentive/2021-22/32439), financial support through the Core Research Grant of SERB, New
Delhi (CRG/2021/000907) and thanks the Inter-University Centre for Astronomy and Astrophysics (IUCAA), Pune for associateship.
\begin{acknowledgments}

\end{acknowledgments}

%

\vspace{5mm}




\bibliography{sample631}{}
\bibliographystyle{aasjournal}

\appendix
\renewcommand{\thefigure}{A\arabic{figure}}

\setcounter{figure}{0}

 \section{Theoretical Framework} 
\label{app:app1}

\begin{figure}
    \centering
    \includegraphics[width=0.8\linewidth]{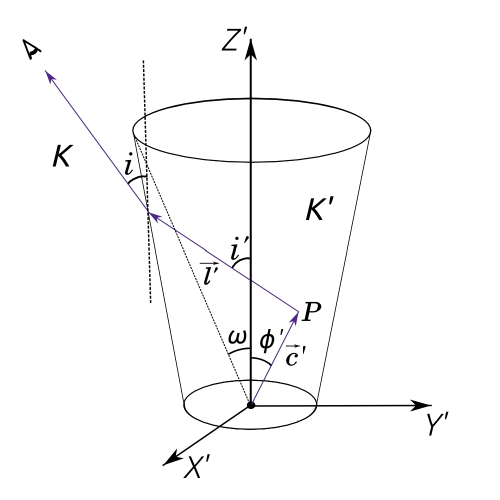}
    \caption{Schematic representation of the relativistic funnel geometry and the relevant coordinate frames. The central primary X-ray source is located at the origin, obscured by the surrounding optically thick material. The scattering medium is confined to a funnel with a half-opening angle $\omega$ and propagates radially outward along the symmetry axis ($Z$-axis) with a macroscopic bulk velocity $v$ (where $\beta = v/c$). The laboratory frame ($K$) represents the rest frame of the central source and the distant observer, who views the system at an inclination angle $i$. The comoving frame ($K^{\prime}$) represents the rest frame of the outflowing plasma.}
    \label{fig:fig5}
\end{figure}
 
 To construct the relativistic scattering model, we use the radiative transfer equations to account for the motion of the scattering medium. We transition from a Static Euclidean framework to a Special Relativistic framework as described in Section~\ref{sec:theoretical framework}.
 \subsection{Frames of Reference}
Two inertial frames of reference are introduced for the derivation as described in the main text. The laboratory frame ($K$) corresponds to the rest frame of the observer and the central compact source. The comoving frame ($K'$) is the rest frame of the plasma element involved in the scattering process. The funnel geometry is illustrated in Figure~\ref{fig:fig5}, where an incident ray from the central source is scattered at point $P$ and propagates toward the observer, located at an inclination $i$ in the laboratory frame ($K$) and $i'$ in the comoving frame ($K'$). 

 \subsection{The Relativistic Doppler Factors and aberration}
 A fundamental theorem of radiative transfer is the Lorentz invariance of the quantity $I_{\nu}/\nu^3$.
\begin{equation}
\frac{I_{\nu}}{\nu^3} = \frac{I'}{(\nu')^3}
\end{equation}
Consider a flux $F_p$ from the central source incident on a scattering point ($P$) inside the funnel. Let $\vec{c}$ and $\vec{c}'$ denote the vectors connecting the central source to the scattering point $P$ in the lab frame ($K$) and comoving frame ($K'$), respectively, with magnitudes $c = |\vec{c}|$ and $c' = |\vec{c}'|$. Similarly, after scattering at $P$, the photon propagates along the directions $\vec{l}$ and $\vec{l}'$ in the frames $K$ and $K'$, with corresponding path lengths $l = |\vec{l}|$ and $l' = |\vec{l}'|$ before escaping from the funnel medium. \\
In the comoving frame ($K'$), this flux is attenuated due to absorption along the incident photon path from the central source to the scattering point $P$, characterized by an optical depth $\tau_c'$ given by:
\begin{equation}
\tau_c' = \tau_c = \alpha'\, c',
\end{equation}
where the optical depth $\tau_c = \int \alpha\, dl$ is Lorentz invariant and $\alpha'$ is the corresponding absorption coefficient of the medium.

The effective flux seen by the electron in the rest frame of the scattering electron, $F'_p$, is:
\begin{equation}
F'_p = F_p(Lab) \cdot D_{in}^3 = I_0 \frac{\pi \Gamma^2}{c'^2} e^{-\alpha' c'} D_{in}^3
\end{equation}
where $I_{0}$ is the initial intensity and $\Gamma$ is the radius of the central source respectively. \\
 The electrons oscillate in response to the incident field $F'_p$. The amount of energy scattered per unit time, per unit volume, per unit solid angle is given by the emissivity $j'$.
 \begin{equation}
 j' = n F'_p \frac{d\sigma'}{d\Omega'}
 \label{eq:eqa1}
 \end{equation}
 where $n$ is the electron number density and $\frac{d\sigma'}{d\Omega'}$ is the differential Thomson cross-section in the rest frame of Electron. 
 For unpolarized incident light:
 \begin{equation}
 \frac{d\sigma'}{d\Omega'} = \frac{3\sigma_T}{16\pi} (1 + \cos^2 \Psi')
 \end{equation}
 
 where $\Psi'$ is the scattering angle in the rest frame of Electron and $\sigma_T$ is the Thomson scattering cross-section. Substituting 
 $F'_p$ in Equation~\ref{eq:eqa1} and using Lorentz invariance of optical thickness, $\tau_{c} = \tau_{c}'$:
 \begin{equation}  
 j' = n I_0 \frac{\pi \Gamma^2}{c'^2} e^{-\tau_{c}} \frac{3\sigma_T}{16\pi} (1 + \cos^2 \Psi')D_{in}^3
 \end{equation}
 \\

In the comoving frame ($K'$), the light ray travels a total distance $l'+c'$ in the volume of funnel before coming out from the scattering medium.

We can write the intensity relation for scattered radiation which leaves the volume element with intensity $dI_{s}'$ as:
\begin{equation}
dI_{s}' = n F'_p \frac{d\sigma'}{d\Omega'} e^{-\tau_{l}}   dl'
\end{equation}

 where $dl'$ is the path length through the element in $K'$. Now, we transform this scattered intensity back to the Lab frame to determine what the observer sees. Using the intensity transformation:
 \begin{equation}
 dI_{obs} = dI_{s}' \cdot D_{out}^3
 \end{equation}
 Substituting the expression for $dI_{s}'$:
 \begin{equation}
 dI_{obs} = \left( n I_0 \frac{\pi \Gamma^2}{c'^2} e^{-(\tau_{c}+\tau_{l})} D_{in}^3 \frac{d\sigma'}{d\Omega'} \right) dl' D_{out}^3
 \end{equation}
 \\
 To find the total scattered flux $F_s$, we integrate the intensity contribution $dI_{obs}$ over the solid angle of the source and the volume of the funnel.
 \\
 The differential flux $dF_s$ is related to intensity by the solid angle subtended by the scattering element at the observer distance $d$:
 
 \begin{equation}
 dF_s = dI_{obs} d\Omega_{obs} = dI_{obs} \frac{dA_{\perp}}{d^2}
 \end{equation}
 where $dA_{\perp}$ is the infinitesimal area
perpendicular to the line of sight of the observer. Substituting the expression for $dI_{\rm obs}$, the scattered flux received by the observer can be written as:

\begin{equation}
dF_s = \left( n I_0 \frac{\pi \Gamma^2}{c'^2} e^{-(\tau_{c}+\tau_{l})} D_{in}^3 \frac{d\sigma'}{d\Omega'} \right) dl' D_{out}^3 \frac{dA_{\perp}}{d^2}
\end{equation}

We use the Lorentz invariance of optical thickness, 
\begin{equation}
\alpha dl = \alpha' dl',
\end{equation}
which gives
\begin{equation}
dl' = \frac{\alpha}{\alpha'} dl.
\end{equation}

Substituting this transformation, we obtain:
\begin{equation}
dF_s = \left( n I_0 \frac{\pi \Gamma^2}{c'^2} e^{-(\tau_{c}+\tau_{l})} D_{in}^3 \frac{d\sigma'}{d\Omega'} \right) \frac{\alpha dl}{\alpha'} D_{out}^3 \frac{dA_{\perp}}{d^2}
\end{equation}

Using the Thomson differential cross-section for unpolarized radiation:
\begin{equation}
dF_s = D_{in}^3 D_{out}^3 \left( \frac{3 \lambda F_{i,\mathrm{rel}}}{16 \pi} \frac{\alpha'}{c'^2} e^{-(\tau_{c}+\tau_{l})}  (1 + \cos^2 \Psi') \right)  dV
\end{equation}

Where $dV = dldA_{\perp}$ is the infinitesimal volume element in the frame $K$ and $F_{i,\mathrm{rel}}$ denotes the intrinsic flux of the source that would be observed in the absence of any funnel geometry.\\ 
Using cylindrical coordinates, $dV = r\, dr\, d\theta\, dz$, we write:

\begin{equation}
dF_s = D_{in}^3 D_{out}^3 \left( \frac{3 \lambda F_{i,\mathrm{rel}}}{16 \pi} \frac{\alpha'}{c'^2} e^{-(\tau_{c}+\tau_{l})}  (1 + \cos^2 \Psi') \right)  r dr d\theta dz
\end{equation}

We further transform the lengths to the optical thickness by using the following transformations:
\begin{equation}
\tau_{r} = \alpha r \quad and \quad \tau_{z} = \alpha z
\end{equation}

\begin{equation}
dF_s =  \frac{3 \lambda F_{i,\mathrm{rel}}}{16 \pi} \frac{D_{in}^3 D_{out}^3}{\alpha^2 c'^2} e^{-(\tau_{c}+\tau_{l})}  (1 + \cos^2 \Psi')   \tau_{r} d\tau_{r} d\theta d\tau_{z}
\end{equation}

Finally, since the photon path length transforms as $c = c'/D_{\rm in}$, we obtain:
\begin{equation}dF_s = D_{in} D_{out}^3 \frac{3 \lambda F_{i,\mathrm{rel}}}{16 \pi} \frac{1}{ \tau_{c}^2} e^{-(\tau_{c}+\tau_{l})}  (1 + \cos^2 \Psi')   \tau_{r} d\tau_{r} d\theta d\tau_{z}
\end{equation}

 Integrating over the funnel geometry:
 \begin{equation}
 F_s = \iiint D_{in} D_{out}^3 \frac{3\lambda}{16\pi} F_{i,\mathrm{rel}} \frac{(1 + \cos^2 \Psi')}{\tau_c^2} \tau_r d\tau_r d\theta d\tau_z
 \end{equation}
 This equation represents the relativistic scattering master equation. It explicitly replaces the static geometric factors with the relativistic product $D_{in} D_{out}^3$, which varies across the funnel volume depending on the local velocity vector relative to the line of sight. \\
 
 \subsection{Derivation of Polarization Degree}
 In the rest frame of the scattering electron ($K'$), the polarization properties of Thomson-scattered radiation depend only on the scattering angle $\Psi'$. The resulting polarization degree is therefore

\begin{equation}
PD'=\frac{1-\cos^{2}\Psi'}{1+\cos^{2}\Psi'} 
\label{eq:eq1}
\end{equation}

The scattered radiation is partially polarized and can be written as the sum of polarized and unpolarized components:
\begin{equation}
I_{ST} = I_{sp} + I_{sup}
\end{equation}

\begin{equation}
I_{ST} = I_s \frac{(1 - \cos^2\Psi')}{(1 + \cos^2\Psi')} + 2I_s \frac{\cos^2\Psi'}{(1+\cos^2\Psi')}
\end{equation}

For a single photon trajectory, the polarization degree is Lorentz invariant provided that the polarization plane is not rotated by propagation effects such as Faraday rotation. This assumption is well justified for X-ray photons in the present context. Consequently, the observed polarization degree in the laboratory frame can be approximated as

\begin{equation}
PD_{\rm obs} \approx PD'
\end{equation}

We follow a similar treatment for the polarization measurement as described by \cite{Chaurasia_2025}. The coordinate system is rotated by an angle $i'$ about the $X'$-axis such that the observer lies along the $Z'$-axis. In this rotated frame, the polarization vector has two components along the $X'_{R}$ and $Y'_{R}$ axes.

Using Malus's law, the intensity of scattered polarized radiation along $X'_{R}$ will be:
\begin{equation}
I_{X'_{RP}} = I_{s}\frac{(1 - \cos^2\Psi')}{(1 + \cos^2\Psi')}   \cos^2\eta'
\end{equation}
where $\eta'$ is the angle between the polarization vector and the $X'_{R}$ axis. 

The intensity along $X'_{R}$ is transformed into the observer's frame along $X_{R}$. We compute the observed flux along $X_{R}$ and $Y_{R}$ using the same procedure as described above, and to derive the polarization degree (PD). \\

The PD of the observed flux will be:
\begin{equation}
PD = \frac{I_{max}-I_{min}}{I_{max}+I_{min}} = \frac{F_{X_{RP}}- F_{Y_{RP}}}{F_{s}}
\end{equation}

The above expression is used to evaluate the polarization degree of the observed radiation.

\section{Narrow-Funnel Limit}
\label{app:app3}
We consider the case of an infinitely narrow cylinder for example and apply this model to the case of Cygnus X-3 to demonstrate how a single geometric configuration can account for the polarization observed in different spectral states. The inclination of the system is taken to be $i=30^{\circ}$ ($\cos i=0.866$), while the funnel geometry is assumed to remain unchanged between states.

During the soft state the radio jet is suppressed and the outflow is likely dominated by a slow, optically thick wind or by quasi-static funnel walls. We therefore consider the limit $\beta \approx 0$. In this case,

\begin{equation}
\cos \Psi'=\frac{\cos 30^{\circ}-0}{1-0}=0.866 ,
\end{equation}

which gives

\begin{equation}
PD_{\rm static}= 0.143 .
\end{equation}

Thus, the model predicts a polarization degree of $\sim14\%$, which is consistent with the observed value of $\sim12\%$ once geometric depolarization effects are taken into account. In particular, the finite opening angle of the funnel leads to averaging over multiple scattering directions, slightly reducing the net polarization compared to the idealized single-ray prediction. These results indicate that the soft spectral state of Cygnus X-3 corresponds to the static limit of the funnel-scattering model.

In the hard spectral state the system becomes radio-loud, indicating the presence of an active jet. In this regime the scattering medium, which may correspond to the jet sheath or an entrained disk wind, is expected to move with a mildly relativistic velocity. We therefore consider a finite bulk velocity $\beta$ and determine the value required to reproduce the observed polarization degree of $\sim23\%$.

Setting $PD=0.23$ in Eq.~\ref{eq:eq1}, we obtain

\begin{equation}
\frac{1-\mu'^2}{1+\mu'^2}=0.23 
\end{equation}

where $\mu'=\cos\Psi'$. Solving for $\mu'$ gives

\begin{equation}
\mu'^2 = 0.626 \Rightarrow \cos\Psi' \approx 0.79 
\end{equation}

Using the relativistic aberration relation

\begin{equation}
\cos\Psi'=\frac{\cos i-\beta}{1-\beta\cos i}
\end{equation}

and adopting $i=30^{\circ}$ ($\cos i=0.866$), we solve for $\beta$:

\begin{equation}
\frac{0.866-\beta}{1-0.866\beta}=0.79 .
\end{equation}

This yields

\begin{equation}
\beta \approx 0.24 .
\end{equation}

Thus, a modest increase in the bulk velocity of the scattering medium to $\beta \approx 0.24$ naturally raises the polarization degree from $\sim14\%$ to $\sim23\%$. Such velocities ($\sim0.3$-$0.4,c$) are consistent with the speeds inferred for winds and jet sheaths in microquasar systems (\citealt{10.1111/j.1745-3933.2010.00834.x}).

If the bulk velocity increases further, the model predicts that even higher polarization degrees are possible. The maximum polarization occurs when the scattering angle in the comoving frame reaches $\Psi'=90^{\circ}$, corresponding to $\cos\Psi'=0$. From the aberration relation this condition implies

\begin{equation}
\beta=\cos i .
\end{equation}

For $i=30^{\circ}$ this gives $\beta \approx 0.866$. In this limit the observer receives photons that were scattered at $90^{\circ}$ in the comoving frame, and the polarization degree approaches its theoretical maximum ($PD \rightarrow 1$).

The fact that Cygnus X-3 exhibits a polarization degree of $\sim23\%$ therefore suggests that the scattering region is only mildly relativistic ($\beta \sim 0.3$-$0.4$), rather than ultra-relativistic ($\beta \gtrsim 0.9$).




\end{document}